# Excitonic Resonance Effects and Davydov Splitting in Circularly Polarized Raman Spectra of Few-Layer WSe$_2$

*Sanghun Kim[1], Kangwon Kim[1], Jae-Ung Lee, and Hyeonsik Cheong*

Department of Physics, Sogang University, Seoul 04107, Korea

**E-mail:** hcheong@sogang.ac.kr

**Notes:** [1]These authors contributed equally.

**Abstract**

Few-layer tungsten diselenide (WSe$_2$) is investigated using circularly polarized Raman spectroscopy with up to eight excitation energies. The main $E_{2g}^1$ and $A_{1g}$ modes near 250 cm$^{-1}$ appear as a single peak in the Raman spectrum taken without consideration of polarization but are resolved by using circularly polarized Raman scattering. The resonance behaviors of the $E_{2g}^1$ and $A_{1g}$ modes are examined. Firstly, both the $E_{2g}^1$ and $A_{1g}$ modes are enhanced near resonances with the exciton states. The $A_{1g}$ mode exhibits Davydov splitting for trilayers or thicker near some of the exciton resonances. The low-frequency Raman spectra show shear and breathing modes involving rigid vibrations of the layers and also exhibit strong dependence on the excitation energy. An unidentified peak at ~19 cm$^{-1}$ that does not depend on the number of layers appears near resonance with the B exciton state at 1.96 eV (632.8 nm). The strengths of the intra- and inter-layer interactions are estimated by comparing the mode frequencies and Davydov splitting with the linear chain model, and the contribution of the next-nearest-neighbor interaction to the inter-layer interaction turns out to be about 34% of the nearest-neighbor interaction. Fano resonance is observed for 1.58-eV excitation, and its origin is found to be the interplay between two-phonon scattering and indirect band transition.

1. **Introduction**

Transition metal dichalcogenides (TMDs) are drawing much attention for possible application as optoelectronic devices, especially for flexible devices [1–3]. Raman spectroscopy is extensively used as a fast and non-destructive characterization tool to study the structural properties of TMDs [4,5]. On the other hand, the optical properties of TMDs are heavily modulated by the excitonic effect because of large exciton binding energies on the order of a few hundred meV [6–10]. The excitonic effects influence Raman scattering through various resonance-related effects, resulting in a number of intriguing phenomena [11–20]. Therefore, it is crucial to be able to distinguish Raman features specific to resonance conditions from those that are intrinsic vibrational features. Although TMDs share many similarities such as indirect-to-direct gap transition from bulk to monolayer, strongly coupled valley-spin, and a large exciton binding energy [6–10, 21–24], there exists subtle differences depending on the combination of transition metal and chalcogen elements. For example, the lowest-energy exciton states in $WS_2$ and $WSe_2$ are dark excitons whereas the corresponding states in $MoS_2$ and $MoSe_2$ are bright [8,25]. Similarly, resonance Raman behaviors of the TMDs have many common features such as strong resonance and activation of forbidden modes at some exciton states, which depends on the compound. For example, the strong resonance behavior is observed at the A exciton state of $MoS_2$ [12] but in $MoSe_2$, the strongest resonance is observed at the C exciton state [17,18]. Furthermore, Davydov splitting of the intralayer vibration mode due to interlayer interaction has been reported for $MoSe_2$, $MoTe_2$, and $WS_2$ only [15,17,19,26–30]. Since $WSe_2$ has a relatively small exciton transition energy (~1.65 eV for monolayer) compared with other TMDs, it is often used to form heterostructures with other TMDs [31–33]. Therefore, it is important to establish the basic physical properties of this material. Furthermore, comparing different resonance behaviors of similar TMD materials might shed light on the origin of novel phenomena observed in these materials.

Monolayer $WSe_2$ comprises a tungsten layer sandwiched between selenium layers, covalently bonded to one another to form a trilayer (TL), in a hexagonal 1H structure [34]. Each TL is bonded by relatively weak van der Waals interaction with neighboring TLs and is stacked in such a way that the selenium atoms on the second layer sit right on top of the tungsten atoms on the first layer to form a so-called 2H structure [35], which is the most stable polytype [34,36]. Figure 1(a) shows the crystal structure of 2H-$WSe_2$. Due to the weak interaction between the TLs, it can be easily cleaved. Monolayer (1TL) $WSe_2$ is a direct band gap semiconductor with the lowest energy optical transition at 1.65 eV near the $K$ or $K'$ point of the Brillouin zone. For 2TL or thicker, the band structure is indirect, but the direct-gap photoluminescence is still visible [37,38]. The optical absorption spectrum of $WSe_2$ is dominated by four peaks: at ~1.65, ~2.1, ~2.4, and ~2.9 eV for 1TL. Some research groups identified them as A, B, A′, and B′ excitons, where the A exciton is formed between the conduction band minima and the top of the valence band at the $K$ or $K'$ points and the B exciton between the conduction band minima and the spin-orbit split valence band. A′ and B′ are described as the excited states of A and B excitons, respectively [11,37,39–41]. However, other groups assigned them to four independent excitonic states of A, B, C, and D excitons [42–44], where the C exciton is formed between states in the *KΓ* direction of the Brillouin zone, and the D exciton originates from the same bands as the B exciton along the *KM* direction. The origins of these excitonic states are neither well established yet nor the focus of this work. Nevertheless, we will use the notation of A, B, C, and D excitons since there does not seem to be an obvious correlation between A and B excitons and A′ and B′ excitons. The A, C, and D peaks in the absorption spectrum redshift with increasing number of layers, whereas the B peak does not shift appreciably [37]. The resonance effects, when the excitation laser energy is close to these excitonic peak energies, are systematically studied in this work.

Davydov splitting in TMDs is lifting of the degeneracy in the *intralayer* vibrational modes due to *interlayer* interactions and has been observed in $MoTe_2$, $MoSe_2$, and $WS_2$ [15,17–19,26–30]. Since the splitting directly correlates with the interlayer coupling, it is a good indicator of interlayer interaction in layered materials. Furthermore, Davydov splitting affects the Raman selection rule of the mode so as to make some forbidden modes become allowed in few-layer cases. It has been found that the relative intensities of the split peaks change drastically depending on the layer number and the excitation laser energy, usually near excitonic resonances [15,17,29]. Therefore, it is convenient to combine the study of the resonance effects with that of Davydov splitting. In this work, we studied the Raman spectra of 1 to 8 TL $WSe_2$ using 8 different excitation energies to study the resonance effects and Davydov splitting. In order to resolve the main intralayer modes with similar frequencies, we employ circularly polarized Raman scattering. From the measured Raman mode frequencies and Davydov splitting, the force constants of interlayer and intralayer interactions are estimated.

2. Method

Few-layer $WSe_2$ samples were prepared by mechanically exfoliating on Si substrates with a 285-nm $SiO_2$ layer from single-crystal bulk $WSe_2$ flakes (HQ Graphene) (see supplementary information for representative optical microscope images). The number of TLs was confirmed by combination of optical contrast, Raman, and PL measurements (see supplementary information for the PL spectra of few-layer WSe$_2$ [26,37,38,45]. Micro-Raman measurements were performed with eight different excitation sources: the 325 and 441.6 nm (3.82 and 2.81 eV) lines of a He-Cd laser; the 457.9, 488, and 514.5 nm (2.71, 2.54, and 2.41 eV, respectively) lines of an Ar$^+$ laser; the 532 nm (2.33 eV) line of a diode-pumped-solid-state (DPSS) laser; the 632.8 nm (1.96 eV) line of a He-Ne laser; and the 784.8 nm (1.58 eV) line of a diode laser. The laser beam was focused onto a sample by a 50× microscope objective lens (0.8 N.A.) for all excitation wavelengths except for the 325 nm excitation for which a 40× UV objective lens (0.5 N.A.) was used. The scattered light was collected and collimated by the same objective and dispersed by a Jobin-Yvon Horiba iHR550 spectrometer (1200 grooves/mm for 784.8 nm and 2400 grooves/mm for all the other excitation wavelengths). For detecting the signal, a liquid-nitrogen-cooled back-illuminated charge-coupled-device (CCD) detector was used. To access the low-frequency range below 50 cm$^{-1}$, volume holographic filters (Ondax and OptiGrate) were used to clean the laser lines and reject the Rayleigh-scattered light. The laser power was kept below 50 μW in order to avoid local heating of the sample. The spectral resolution ranged between 0.3 cm$^{-1}$ (1.96 eV) and 1.5 cm$^{-1}$ (3.82 eV). In order to extract the intrinsic resonance effects, the Raman intensities were calibrated by the intensity of the Si Raman peak at 520 cm$^{-1}$ for each excitation energy to correct for the efficiency of the detection system. In addition, the interference effects due to multiple reflection from the substrate [46] and the resonance Raman effects of Si [47] are considered. The wavelength dependent refractive indices are taken from the measured values for 1TL WSe$_2$ [42]. Because of the limitations of available data, bulk values [48] were used for the refractive indices in the UV region (for the excitation wavelength of 325 nm). The detailed calibration procedure has been published elsewhere [12]. Figure 1(b) shows the schematic of the circularly polarized Raman measurement system. For the circularly polarized light, we use a polarizer and a quarter wave plate before the objective lens. The scattered light from the sample passes through the same quarter-wave plate, becoming linearly polarized in parallel or orthogonal direction depending on the

## 3. Results and Discussion

Figure 1(c) is an (unpolarized) Raman spectrum of 4TL WSe$_2$ measured with the excitation energy of 2.54 eV (488 nm). The peaks marked by '*' are due to Brillouin scattering from the Si substrate. In the low-frequency region (<50 cm$^{-1}$), there are shear (blue circles) and breathing (red triangles) modes, which correspond to the in-plane and out-of-plane direction interlayer vibrations, respectively [45,49]. In the high frequency region (100–400 cm$^{-1}$), four first order Raman modes are observed: E$_{1g}$ (~175 cm$^{-1}$), E$_{2g}^1$ (~248 cm$^{-1}$), A$_{1g}$ (~251 cm$^{-1}$), and A$_{2u}$ (~308 cm$^{-1}$) modes. Since the point symmetry groups of bulk and few-layer WSe$_2$ are $D_{6h}$ for bulk, $D_{3h}$ for all odd TLs, and $D_{3d}$ for all even TLs, the mode notations should be different depending on the number of TLs even though they originate from similar vibrational modes [28,35,45,49]. For simplicity, however, we will use the bulk notations in the following discussion regardless of the number of TLs. The E$_{1g}$ mode is Raman active but forbidden in backscattering for bulk and odd-TL WSe$_2$. However, a small signal from this mode is observed regardless of the number of TLs except for the monolayer case (see supplementary information figures S2-S4) because a significant portion of the incident and scattered light is not normal to the sample plane due to the large numerical aperture. The A$_{2u}$ mode is infrared active and Raman inactive, but for few-layer WSe$_2$, there is a small Davydov splitting of this mode due to interlayer interaction. Some of the split modes are Raman active, and therefore there is a finite Raman signal in the vicinity of the A$_{2u}$ mode. Since the splitting is very small, the Davydov split peak appears at the same frequency as the main A$_{2u}$ peak. Other weak peaks are also observed. The weak signals at ~220 cm$^{-1}$ and ~268 cm$^{-1}$ have been assigned to the E-mode at the *K*-point, E(*K*), and the A mode at the *M*-point, A(*M*), respectively [11,38]. In addition, the signal near ~260 cm$^{-1}$ is ascribed to 2-phonon scattering of the longitudinal acoustic phonon at the *M* point [2LA(*M*)], the peaks at 360 cm$^{-1}$ and 373 cm$^{-1}$ to the sum of LO and TO phonon of E$_{2g}^1$ mode and LA at the $M$-point [E$_{2g}^1$,LO($M$)+LA($M$) and E$_{2g}^1$,TO($M$)+LA($M$)], respectively, and the signal at ~393 cm$^{-1}$ to 3LA($M$) [11,38,50].

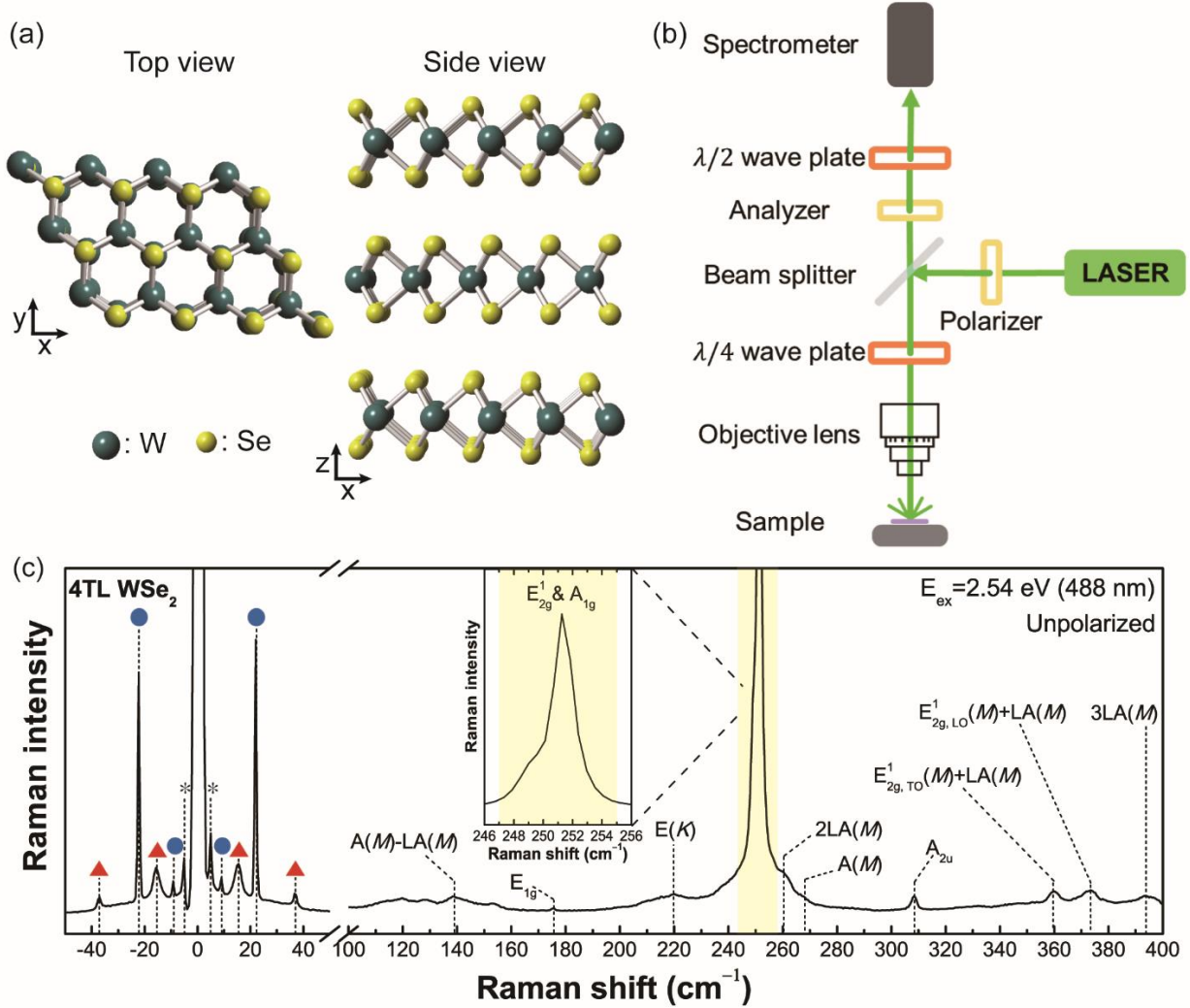

**Figure 1.** (a) Crystal structure of 2H-WSe$_2$. (b) Schematic of the circularly polarized Raman measurement system. (c) Unpolarized Raman spectrum of 4TL WSe$_2$ measured with excitation energy of 2.54 eV (488 nm).

The E$_{2g}^1$ and A$_{1g}$ modes have similar frequencies and so are not resolved. In order to resolve these two peaks, circularly polarized Raman scattering is utilized [51]. If $\sigma_i$ and $\sigma_o$ represent the circular polarizations of the incident and outgoing photons, respectively, the Raman cross section is proportional to $|\langle \sigma_o|R|\sigma_i \rangle|^2$, where $R$ represents the Raman tensor of the mode. Circularly polarized light propagating in the z direction is represented by $\sigma\pm = \frac{1}{\sqrt{2}}\begin{pmatrix} 1 \\ \mp i \\ 0 \end{pmatrix}$. The Raman tensor of the $A_{1g}$ mode is given by $R = \begin{pmatrix} a & 0 & 0 \\ 0 & a & 0 \\ 0 & 0 & b \end{pmatrix}$ for all TLs. For the case of the same circular polarization in backscattering geometry [$(\sigma+\sigma+)$ or $(\sigma-\sigma-)$], the intensity of the $A_{1g}$ mode is proportional to $|\langle \sigma_o|R|\sigma_i \rangle|^2 = a^2$.

On the other hand, if $\sigma_o$ and $\sigma_i$ are opposite [$(\sigma+\sigma-)$ or $(\sigma-\sigma+)$],

$|\langle \sigma_o|R|\sigma_i \rangle|^2 = |\langle \sigma+|R|\sigma- \rangle|^2 = 0$. Therefore, the $A_{1g}$ mode is observed only in the same circular polarization configurations. Meanwhile, the Raman tensors of the $E_{2g}^1$ modes are given by

$R = \begin{pmatrix} 0 & -d & 0 \\ -d & 0 & 0 \\ 0 & 0 & 0 \end{pmatrix}$ or $R = \begin{pmatrix} d & 0 & 0 \\ 0 & -d & 0 \\ 0 & 0 & 0 \end{pmatrix}$ for bulk or odd TLs and $R = \begin{pmatrix} 0 & -c & -d \\ -c & 0 & 0 \\ -d & 0 & 0 \end{pmatrix}$ or

$R = \begin{pmatrix} c & 0 & 0 \\ 0 & -c & d \\ 0 & d & 0 \end{pmatrix}$ for even TLs. In either case, $|\langle \sigma+|R|\sigma+ \rangle|^2 = |\langle \sigma-|R|\sigma- \rangle|^2 = 0$, whereas

$|\langle \sigma+|R|\sigma- \rangle|^2 = |\langle \sigma-|R|\sigma+ \rangle|^2 = d^2$ or $c^2$. Therefore, the $E_{2g}^1$ modes are observed only in the opposite circular polarization configurations. By combining these results, we conclude that only the $A_{1g}$ mode is observed in the same polarization configuration and the $E_{2g}^1$ mode in the opposite polarization configurations. By using circularly polarized Raman scattering, we were able to resolve the $A_{1g}$ and $E_{2g}^1$ peaks and study their resonance profiles separately. We should note that in Raman scattering with linearly polarized light, only the $E_{2g}^1$ mode is observed in cross polarization and both the $A_{1g}$ and the $E_{2g}^1$ modes are observed in parallel polarization. Zhao *et al*. used this method to identify the two modes for only one excitation energy (1.96 eV) [50]. However, since some of the Davydov-split peaks overlap with the $E_{2g}^1$ mode, Davydov splitting cannot be studied in detail if one used linear polarization.

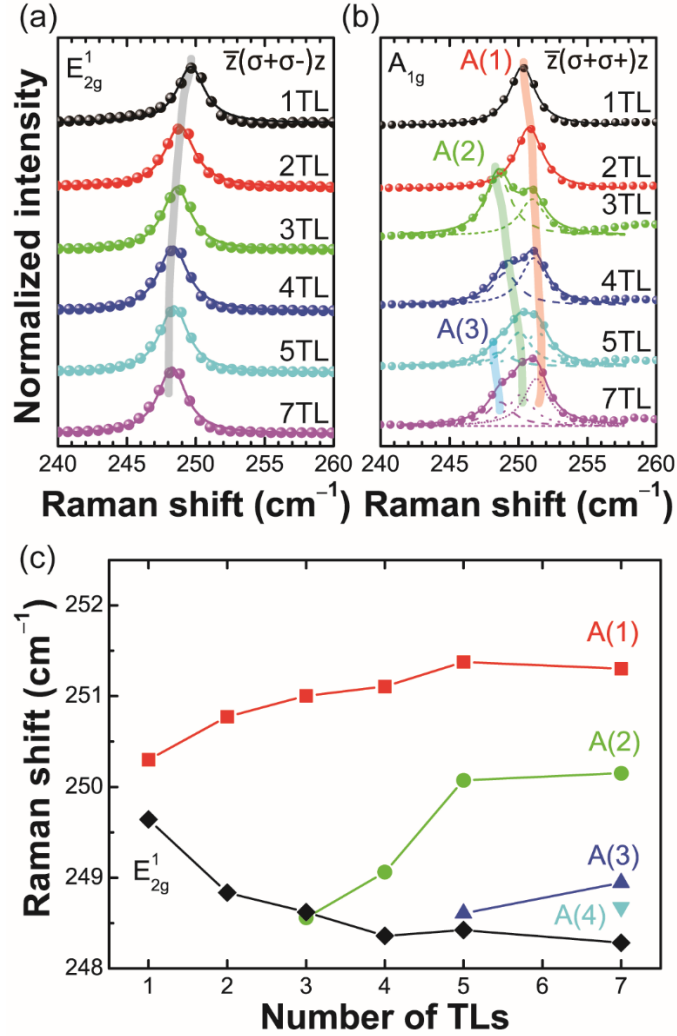

**Figure 2.** Circularly polarized Raman spectra of (a) the $E_{2g}^1$ mode and (b) $A_{1g}$ mode measured with excitation energy of 2.81 eV (441.6 nm). Davydov-split peaks of the $A_{1g}$ mode are resolved. (c) Peak positions as a function of the number of TLs.

Figures 2(a) and (b) show the $E_{2g}^1$ and $A_{1g}$ modes separately, measured with circularly polarized light. The spectra are normalized in order to show the peak shift clearly. The $E_{2g}^1$ and $A_{1g}$ modes, which correspond to in-plane vibration of transition metal and chalcogen atoms and out-of-plane symmetric vibration of chalcogen atoms, respectively, show opposite shifts with the number of TLs: the $E_{2g}^1$ mode redshifts whereas the $A_{1g}$ mode blueshifts as the number of TLs increases. The separation between these two modes can be used to identify the number of TLs as in the case of other TMDs [17,28,50,52].

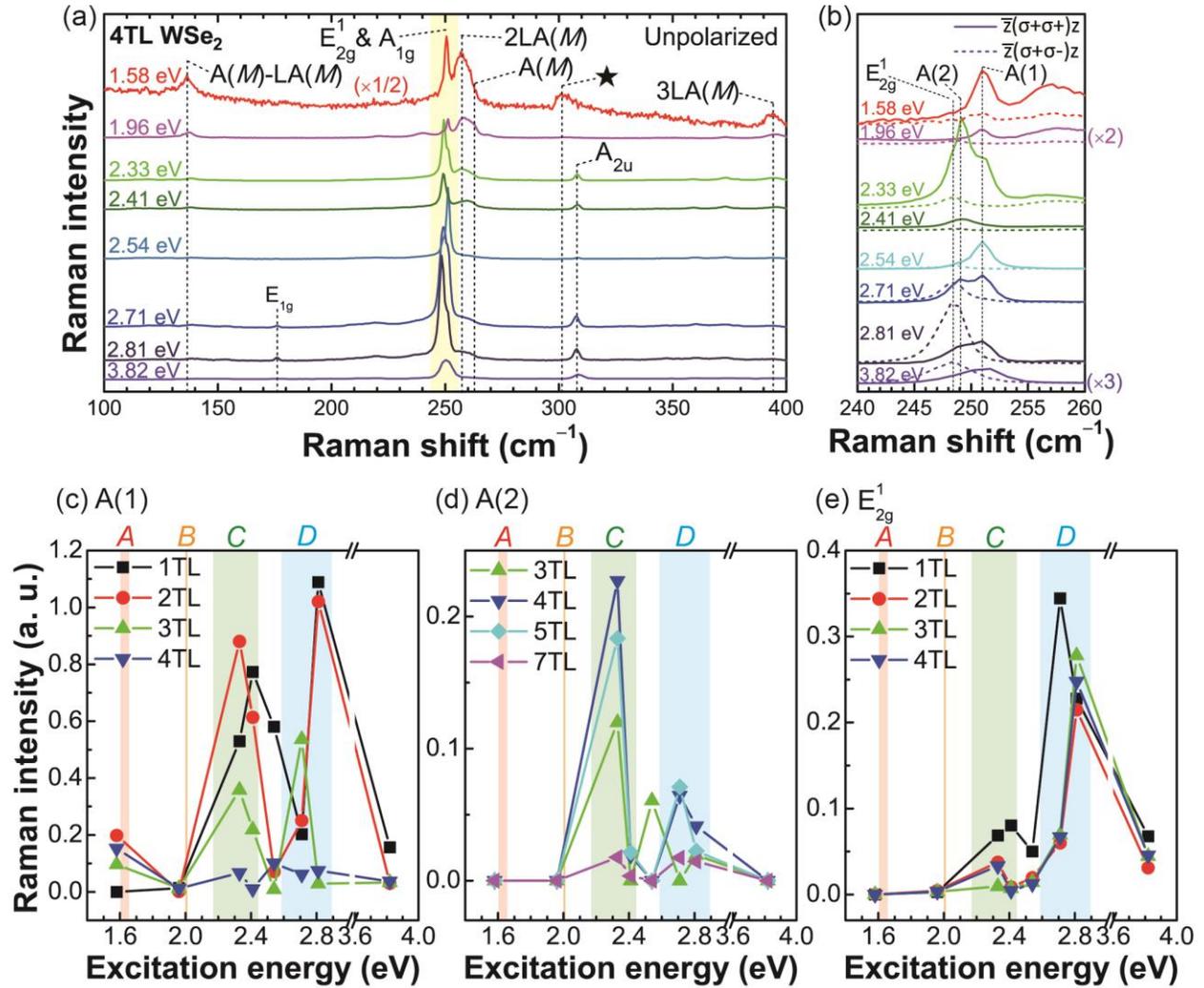

**Figure 3.** Raman spectra of the 4TL WSe$_2$ using eight excitation energies with (a) polarization-unresolved and (b) circularly polarized light. Raman intensity as a function of excitation energies for (c) the A(1), (d) A(2), and (e) $E_{2g}^1$ mode.

Figure 3(a) shows the Raman spectra of 4TL WSe$_2$ measured with 8 different excitation energies without polarization optical components (see supplementary information figures S2-S4 for other

TLs and figures S5-S7 for TL dependence of the spectrum for each excitation energy). The line shape of the unresolved peak of main $E_{2g}^1$ and $A_{1g}$ modes varies greatly depending on the excitation energy, reaffirming the importance of resolving the peaks using circular polarization. The circularly polarized Raman spectra in figure 3(b) show the resonance patterns more clearly, and we find that for some excitation energies the $A_{1g}$ mode is divided into several peaks due to Davydov splitting [53,54]. Similar splitting has been observed in MoTe$_2$, MoSe$_2$, and WS$_2$ [15,17,18,26–30]. In general, an intralayer vibration mode of $n$-TL TMD splits into $n/2$ [$(n+1)/2$] Raman active modes and $n/2$ [$(n-1)/2$] infrared modes for even [odd] $n$. Figure 2(b) shows the deconvolution of the $A_{1g}$ peak to obtain the split peaks. For example, there are two Raman active modes originating from the $A_{1g}$ mode of 4TL, which we call A(1) and A(2) modes, in the order of higher frequency. We also observe in figure 3(b) that the relative ratio of the intensities of A(1) and A(2) depends strongly on the excitation energy. For some excitation energies, A(2) is even stronger than A(1). In general, the highest-frequency mode A(1) is always visible, but the lower frequency modes are observed only near resonance excitations [15,17,28]. Figures 3(c-e) show the resonance profiles of the A(1), A(2), and $E_{2g}^1$ peaks, respectively. All three peaks exhibit strong enhancement near the C and D excitons. The A(1) peaks also shows a small enhancement at the A exciton resonance. On the other hand, none of these peaks show enhancement near the B exciton resonance at ~2.1 eV. These results are consistent with previous resonance Raman measurements without considering polarization [11,16]. Since WSe$_2$ has a relatively smaller bandgap among TMDs, resonance behaviors near the C and D excitons can be more easily examined using available laser lines. Upon close inspection, we find subtle yet important differences between A and $E_{2g}^1$ modes. For example, the D resonance of the A(1) peak seems to shift to lower energy as the number of TLs increases. On the other hand, that for the $E_{2g}^1$ peak shifts to slightly higher energy as the number of TL increases. This difference cannot be noticed without resolving the two peaks by using circular polarization. Since the D exciton energy is known to decrease with thickness [37], the resonance profile of the A(1) peak is consistent with being due to resonance with the D exciton state. However, the thickness dependence of the resonance profile of the $E_{2g}^1$ peak suggests that this resonance may have a different origin. For example, this resonance can be due to transitions between the lower valence band and the higher conduction band at $\Lambda$ (midpoint between $K$ and $\Gamma$) or transitions at the $K$ point between the valence band minima and the higher conduction bands. Furthermore, we also observe that the Davydov-split A(2) peak is strongly enhanced for resonance with the C exciton state, which might shed light on the origin of the enhancement of lower-frequency Davydov-split peaks at specific excitation energies. Although Davydov splitting is common to many TMDs, the resonance conditions for it seems to depend on the material. For $WS_2$, Davydov splitting is enhanced near the A exciton resonance at ~2.0 eV [27]. On the other hand, the splitting was observed for all excitation energies greater than 1.8 eV for $MoTe_2$ [15,28,29]. In the case of $MoSe_2$ and $WSe_2$, Davydov splitting is clearly observed near resonance with the C excitons or the band-to-band transition [17,18]. Since the $A_{1g}$ mode in which Davydov splitting is observed involves vibrations of only the chalcogen atoms, the resonance enhancement of Davydov splitting depends on the kind of chalcogen atoms rather than the transition metal. Further studies are needed to develop a consistent theory that can explain these differences.

As noted before, the (forbidden) $E_{1g}$ mode at ~175 cm$^{-1}$ is clearly observed at higher excitation energies except for the monolayer case. We should mention that this peak is not visible in the case of 3.82 eV excitation because we used a smaller NA objective for this excitation energy. The $A_{2u}$ mode at ~308 cm$^{-1}$ is observed for excitation energies greater than 2 eV, but its intensity is conspicuously smaller for the 2.54 eV excitation. We note that at this excitation energy, the Davydov-split peaks of the main $A_{1g}$ mode are also very weak. This confirms our earlier interpretation the $A_{2u}$ mode is visible due to Davydov splitting [17]. Both the 2LA($M$) and A($M$) signals seem to be enhanced for lower excitation energies (see supplementary information figure S8 for complete resonance profiles for 1–4 TL). At 1.96 eV, for example, the combination of these peaks has a higher intensity than the main $E_{2g}^1$ and $A_{1g}$ modes. This enhancement correlates with resonance with the A and B exciton states near 1.65 and 2.1 eV, respectively. Del Corro *et al*. also reported enhancement of these peaks at the B exciton resonance [11,16]. Similar dramatic enhancement of the multi-phonon scattering signal near resonance has been observed for MoS$_2$ near A and B exciton resonances [12,55].

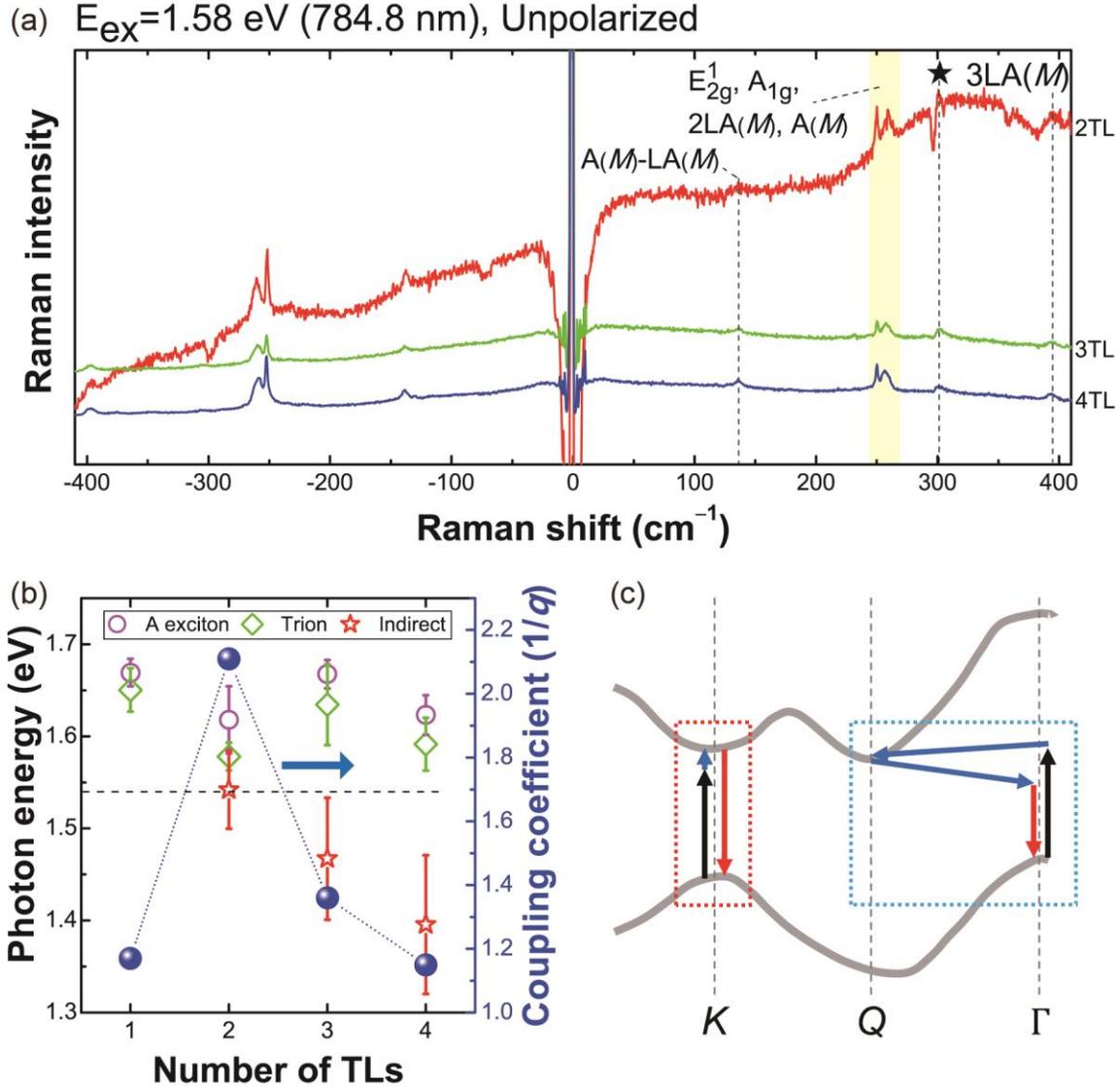

**Figure 4.** (a) Anti-Stokes and Stokes Raman spectra of 2–4 TL WSe$_2$ measured with 1.58 eV excitation energy. (b) Photon energies of A exciton, charged exciton (trion), and indirect gap and coupling coefficient ($1/q$) as a function of the number of TLs. Black dashed line at 1.54 eV shows the scattered photon energy of the BWF line. (c) Schematics of outgoing resonance Raman process at $K$-point (red dotted box) and resonance Raman process from $\Gamma$- to $Q$-point (blue dotted box) responsible for Fano resonance.

For 1.58 eV excitation, we found that the anti-Stokes Raman peaks are stronger than the Stokes Raman peaks for 2TL or thicker (see figure 4(a) and supplementary figure S9), which implies that the anti-Stokes signal is in outgoing resonance. Since the phonon energies of $A_{1g}$ or the $E_{2g}^1$ mode are ~30 meV, the outgoing resonance condition is met if the resonant state has an energy of ~1.61 eV, which in fact is close to the A exciton state for 2TL or thicker. We also observe a Breit-Wigner-Fano (BWF) type [56] signal at ~301 cm$^{-1}$ only for the 1.58 eV excitation regardless of the number of TLs, indicating a Fano-like resonance (see supplementary information for other TLs). This resonance is most prominent for 2TL (see supplementary figure S10). Fano resonance is a quantum interference of a discrete excitation with a continuum and has been observed in TMDs only for a low-frequency shear mode of $WS_2$ [19,20]. The BWF line shape is represented by [57],

$$I(\omega) = I_0 \frac{[1 + 2(\omega - \omega_0)/(q\Gamma)]^2}{[1 + 4(\omega - \omega_0)^2/\Gamma^2]}, \qquad (1)$$

where $\Gamma$ is the broadening parameter, $1/q$ is the coupling constant, and $\omega_0$ is the peak frequency of uncoupled mode. The coupling coefficients $1/q$ obtained from fitting the spectra (see supplementary figure S10) are 1.17, 2.11, 1.36, and 1.15 for 1–4 TL $WSe_2$, respectively, as shown in figure 4(b). Unlike the other phonon modes, the Stokes signal of this BWF line is stronger than the anti-Stokes signal, which indicates that the origin of this Fano resonance is different from the resonance Raman scattering process of the other Raman peaks. In order to analyze Fano resonance, one needs to identify the discrete and continuum excitations responsible for it. For the discrete excitation, phonon emission is a natural candidate. Since there is no phonon mode at this frequency, it should be a combination of two or more phonons. In order to identify the continuum excitation, we examined the PL spectra of 1–4 TL $WSe_2$. The PL signal contains contributions from the exciton, a charged exciton (trion) and the indirect transition (see supplementary figure S11), and figure 4(b) summarizes their energies. For 2TL $WSe_2$, the indirect gap energy (1.54 eV) coincides with the energy of the Raman scattered photon from the BWF line [dashed line in figure 4(b)].

Since Fano resonance is most pronounced for 2TL, this coincidence is an important clue. According to theoretical calculations [58], the indirect transition occurs between the valence band maximum at the $\Gamma$ point and the conduction band minimum at the $Q$ point between $\Gamma$ and $K$ points of the Brillouin zone as shown in figure 4(c). We also find that the position of this peak (301 cm$^{-1}$) is close to the sum of the $E_{1g, LO}$ and ZA phonons at the $Q$-point [$E_{1g, LO}(Q)$+ZA($Q$)] [38]. Based on these observations, we can infer that the Fano resonance is due to the resonance Raman process from $\Gamma$-point of the valence band to $Q$-point the conduction band as shown in blue dotted box in figure 4(c). Since this process is independent of the A exciton state formed at the $K$ point [red dotted box in figure 4(c)], this Fano resonance does not have the same resonance behavior as the other Raman modes that exhibit outgoing resonances. This interpretation is supported by the fact that the coupling coefficient $1/q$ correlates with the proximity of the indirect gap transition energy to the scattered photon energy as demonstrated in figure 4(b).

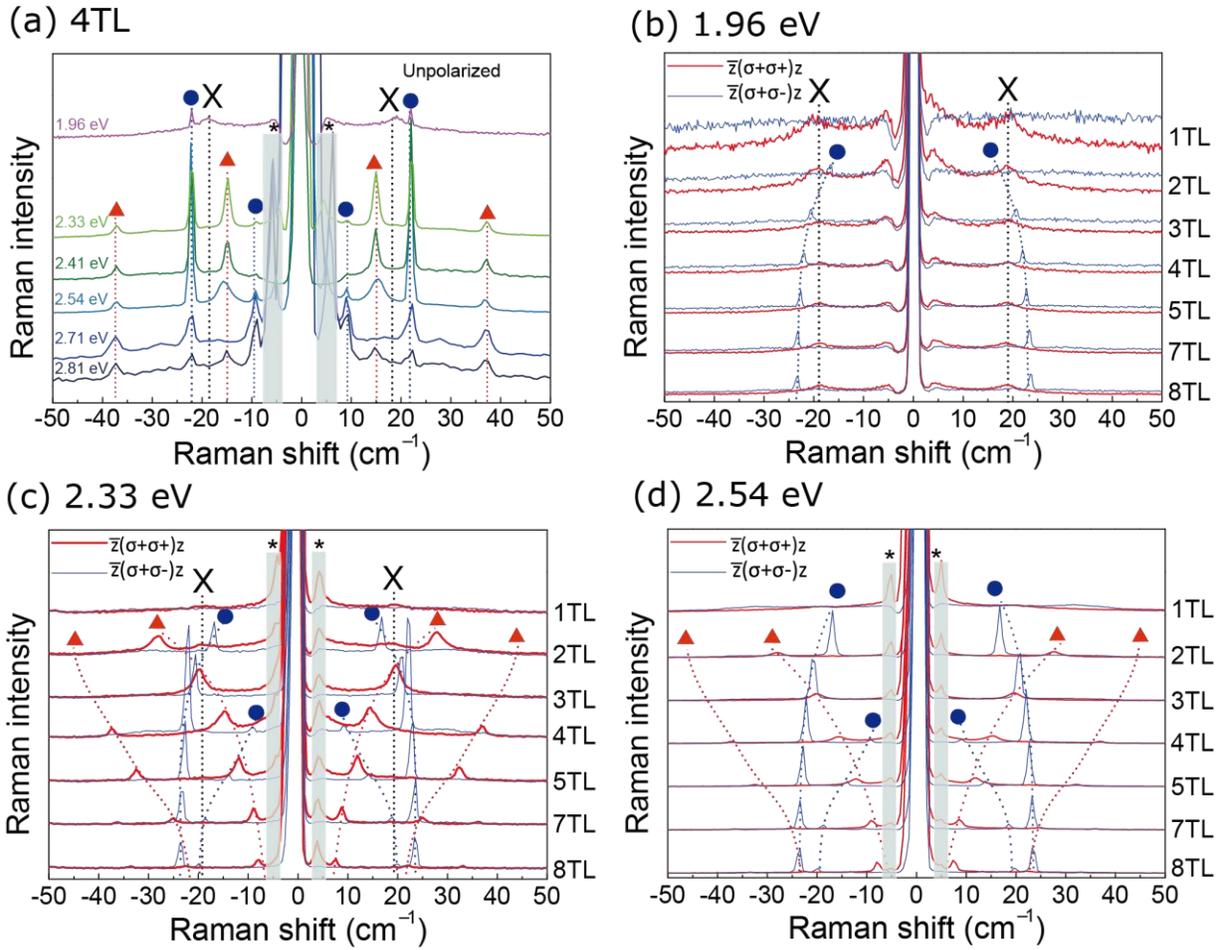

**Figure 5.** (a) Low-frequency Raman spectra of 4TL WSe$_2$ measured with six different laser energies. Shear (●) and breathing (▲) modes are indicated. Dependence of low-frequency circularly-polarized Raman spectrum on number of TLs for excitation energies of (b) 1.96 eV, (c) 2.33 eV, and (d) 2.54 eV. For 1.96 eV and 2.33 eV excitations, an unidentified peak (X) appears for all thicknesses.

Figure 5(a) shows the unpolarized low-frequency Raman spectra of 4TL WSe$_2$ measured with six different excitation energies. The shear and breathing modes are clearly seen, but the relative intensities depend on the excitation energy (see supplementary information figure S8 for resonance profiles of these modes). Figures 4(b-d) show the dependence of the low frequency spectrum on the number of TLs for three excitation energies, respectively. Here, 1.96 eV is close to the B exciton resonance and 2.33 eV to the C exciton resonance, whereas 2.54 eV corresponds to the off-resonance case. Circularly polarized Raman scattering is employed to resolve the breathing (σ+σ+) and the shear (σ+σ−) modes. The strong dependence of the peak positions on the number of TLs can be used as a definitive measure of the number of Ts. In addition, we observe a clear signal at 19 cm$^{-1}$ for the 1.96 eV excitation for all TLs. This peak also appears weakly for the 2.33 eV excitation, and its position does not depend on the number of TLs. Similar low-frequency peaks have been observed in MoS$_2$ and WS$_2$ [12,19,20], but there are important differences. Although two such peaks are observed in WS$_2$ [19,20], there is no indication of a second peak in WSe$_2$. Also, unlike the case of MoS$_2$ where this peak is observed for both same (σ−σ−) and opposite (σ−σ+) circular polarization configurations [12], this peak in WSe$_2$ is observed only for the same circular polarization configuration. Also, the small spin-orbit splitting of the conduction band was suggested as the origin of this peak in MoS$_2$ [59], but the corresponding splitting in WS$_2$ or WSe$_2$ are much larger. For WS$_2$, there has been a suggestion that the two such peaks are from acoustic phonons (TA and LA) with a specific momentum value [19]. This interpretation was based partly on the ratio of two mode frequencies which is close to the ratio of LA/TA mode frequencies, but it was not obvious what determines the specific value of momentum regardless of the number of TLs. At the moment, there is no consistent model that can explain the origin of these peaks, and so we tentatively label it as an 'X' peak. Since it appears only near excitonic resonances, a new kind of collective excitation of the excitons could be the origin, but more studies are needed.

From the positions of the shear and breathing modes, one can calculate the in-plane and out-of-plane interlayer force constants using the linear chain model [45,49]. We obtained the in-plane force constant of $3.06 \times 10^{19}\, Nm^{-3}$ and the out-of-plane force constant of $8.56 \times 10^{19}\, Nm^{-3}$ by considering only the nearest-neighbor interaction between adjacent Se layers. These values are similar to those from previous work [45]. In the case of out-of-plane vibrations, the calculations can be refined by including the Davydov-split peak positions and including the next-nearest-neighbor interactions [17,28] and the surface effect [60]. The detailed calculation procedure is described in Refs. [17] and [28]. By fitting the peak positions of the out-of-plane Raman modes, breathing, A$_{1g}$ including Davydov splitting, and A$_{2u}$ including Davydov splitting (see supplementary information figure S12), we obtained the out-of-plane force constants of WSe$_2$. The fitting results are shown in figure 5, and the obtained parameters are summarized in table 1. The force constant for the nearest-neighbor interlayer interaction (*β*) between Se atoms of the adjacent layer is $4.80\times10^{19}\,Nm^{-3}$, which is only half the value obtained in the simple linear chain model that included only the nearest-neighbor interaction. The force constant for the next-nearest-neighbor interlayer interaction (*γ*) between the Se atoms and the W atoms in the adjacent layer is $1.65\times10^{19}\,Nm^{-3}$, which amounts to about 34 % of the nearest-neighbor interlayer interaction (*β*). This is somewhat larger than those in MoTe$_2$ (~ 20 %) [28] and MoSe$_2$ (~ 30%) [17].

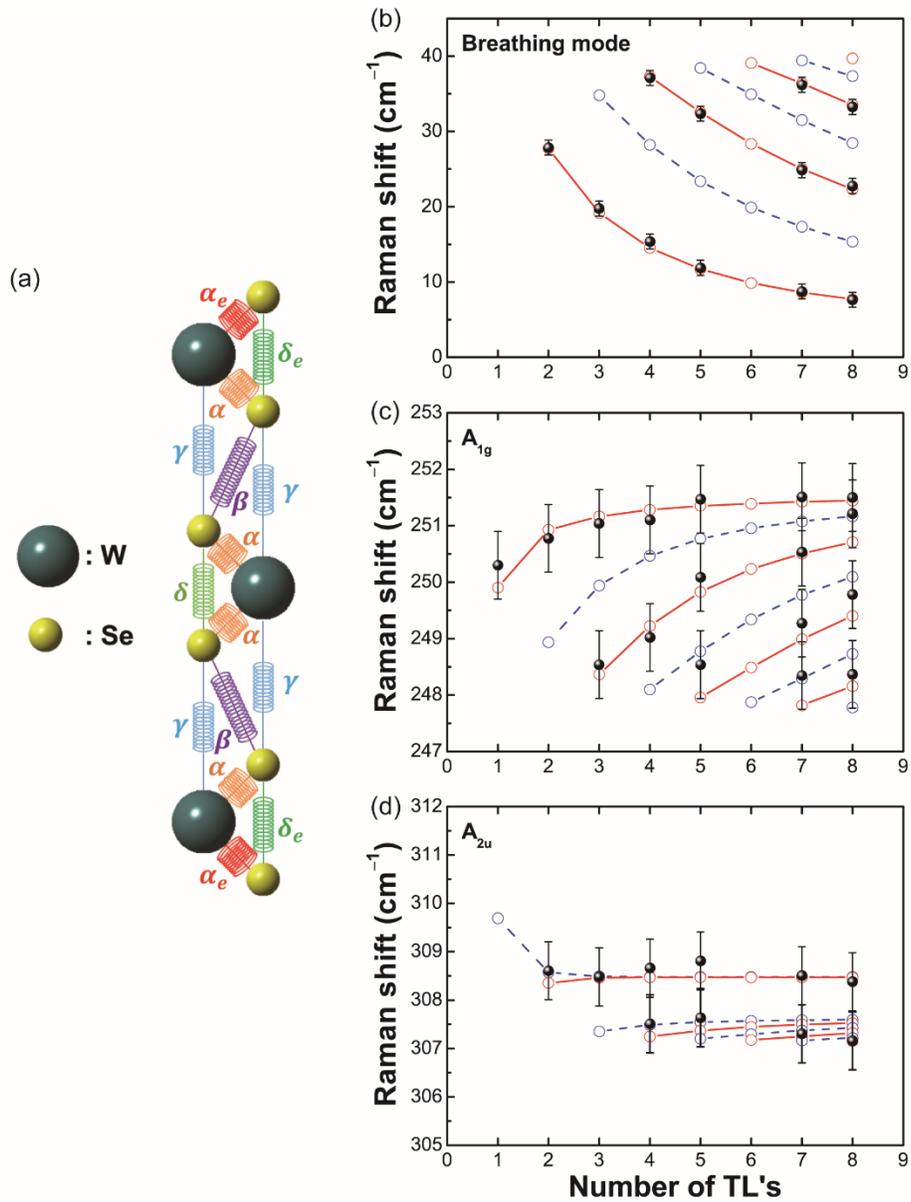

**Figure 6.** (a) Force constants in linear chain model. Out-of-plane Raman modes as a function of the number of TLs: (b) breathing, (c) $A_{1g}$, and (d) $A_{2u}$ modes. Experimental data are represented by filled circles, and open circles indicate calculated values. Red and blue data represent Raman active and inactive modes, respectively.

**Table 1.** Out-of-plane direction force constants per unit area of WSe$_2$ obtained by fitting experimental data to simple linear chain model (LCM) and a model including next-nearest-neighbor interactions.

| | Corresponding interaction | Simple LCM | Including next-nearest-neighbor interaction |
|---|---|---|---|
| $\alpha$ ($10^{19}$ N/m³) | Intra-layer W-Se | 312.5±0.3 | 249.0±0.3 |
| $\alpha_e$ ($10^{19}$ N/m³) | Intra-layer W-Se for surface Se | 310.1±0.3 | 255.3±0.3 |
| $\beta$ ($10^{19}$ N/m³) | Inter-layer Se-Se | 8.56 | 4.80±0.2 |
| $\gamma$ ($10^{19}$ N/m³) | Inter-layer W-Se | | 1.65±0.05 |
| $\delta$ ($10^{19}$ N/m³) | Intra-layer Se-Se | | 26.0±0.4 |
| $\delta_e$ ($10^{19}$ N/m³) | Intra-layer Se-Se for surface Se | | 26.9±0.4 |

## 4. Conclusion

The intralayer $E_{2g}^1$ and $A_{1g}$ modes of few-layer WSe$_2$ near 250 cm$^{-1}$ are resolved by using circularly polarized Raman scattering, and their resonance behaviors are examined for eight excitation energies. Both the modes are enhanced near resonances with the exciton states, and the $A_{1g}$ mode exhibits Davydov splitting for trilayers or thicker near some of the exciton resonances. The low-frequency Raman spectra also show dependence of the shear and breathing mode intensities on the excitation energy. An unidentified peak at ~19 cm$^{-1}$ that does not depend on the number of TLs appears near resonance with the B exciton state at 1.96 eV. The intra- and inter-layer force constants in the out-of-plane direction are estimated by comparing the mode frequencies and Davydov splitting with the linear chain model including the next-nearest-neighbor interaction. The contribution of the next-nearest-neighbor interaction amounts to about 34% of the nearest-neighbor interaction. Fano resonance which depends on the number of TLs is observed, and its origin is found to be the interplay between two-phonon scattering and indirect transition between the $\Gamma$ point of the valence band and the $Q$ point of the conduction band.

**Acknowledgements**

This work was supported by the National Research Foundation (NRF) grant funded by the Korean government (MSIP) (NRF-2016R1A2B3008363 and No. 2017R1A5A1014862, SRC program: vdWMRC center) and by a grant (No. 2011-0031630) from the Center for Advanced Soft Electronics under the Global Frontier Research Program of MSIP.

# Supplementary Information

# Excitonic Resonance Effects and Davydov Splitting in Raman Spectra of Few-Layer WSe$_2$

*Sanghun Kim,*[†] *Kim Kangwon,*[†] *Jae-Ung Lee,*[†] *and Hyeonsik Cheong*[†]

[†]Department of Physics, Sogang University, Seoul 04107, Korea

**Contents:**

- **Figure S1.** Optical microscope images of few-layer WSe$_2$ samples and photoluminescence (PL) spectra of few-layer WSe$_2$.

- **Figure S2.** Raman spectra of 1–3TL WSe$_2$ measured with 8 excitation energies.

- **Figure S3.** Raman spectra of 4–5TL, and 7TL WSe$_2$ measured with 8 excitation energies.

- **Figure S4.** Raman spectra of 8TL and 12 nm WSe$_2$ measured with 8 excitation energies.

- **Figure S5.** Thickness dependence of Raman spectrum of WSe$_2$ measured with excitation energies 1.58 eV, 1.96 eV, and 2.33 eV.

- **Figure S6.** Thickness dependence of Raman spectrum of WSe$_2$ measured with excitation energies 2.41 eV, 2.54 eV, and 2.71 eV.

- **Figure S7.** Thickness dependence of Raman spectrum of WSe$_2$ measured with excitation energies 2.81 eV and 3.82 eV.

- **Figure S8.** Raman intensity as a function of excitation energies for the E$_{1g}$, A$_{2u}$, 2LA(*M*), shear, and breathing modes.

- **Figure S9.** Anti-Stokes and Stokes Raman spectra of few-layer WSe$_2$ measured with 1.58 eV excitation energy.

- **Figure S10.** Breit-Wigner-Fano (BWF) line fitting of Raman spectra near 301 cm$^{-1}$ measured with excitation energy of 1.58 eV.

- **Figure S11.** Photoluminescence (PL) spectra of 1-4 TL WSe$_2$ with deconvolution of A exciton, trion, and indirect gap transition signals.

- **Figure S12.** Excitation energy dependent low-frequency Raman spectra of 1–5TL and 7TL WSe$_2$.

- **Figure S13.** Excitation energy dependent low-frequency Raman spectra of 8TL and 12 nm WSe$_2$.

- **Figure S14.** Raman spectra of the A$_{2u}$ mode measured with excitation energy of 2.81 eV (441.6 nm).

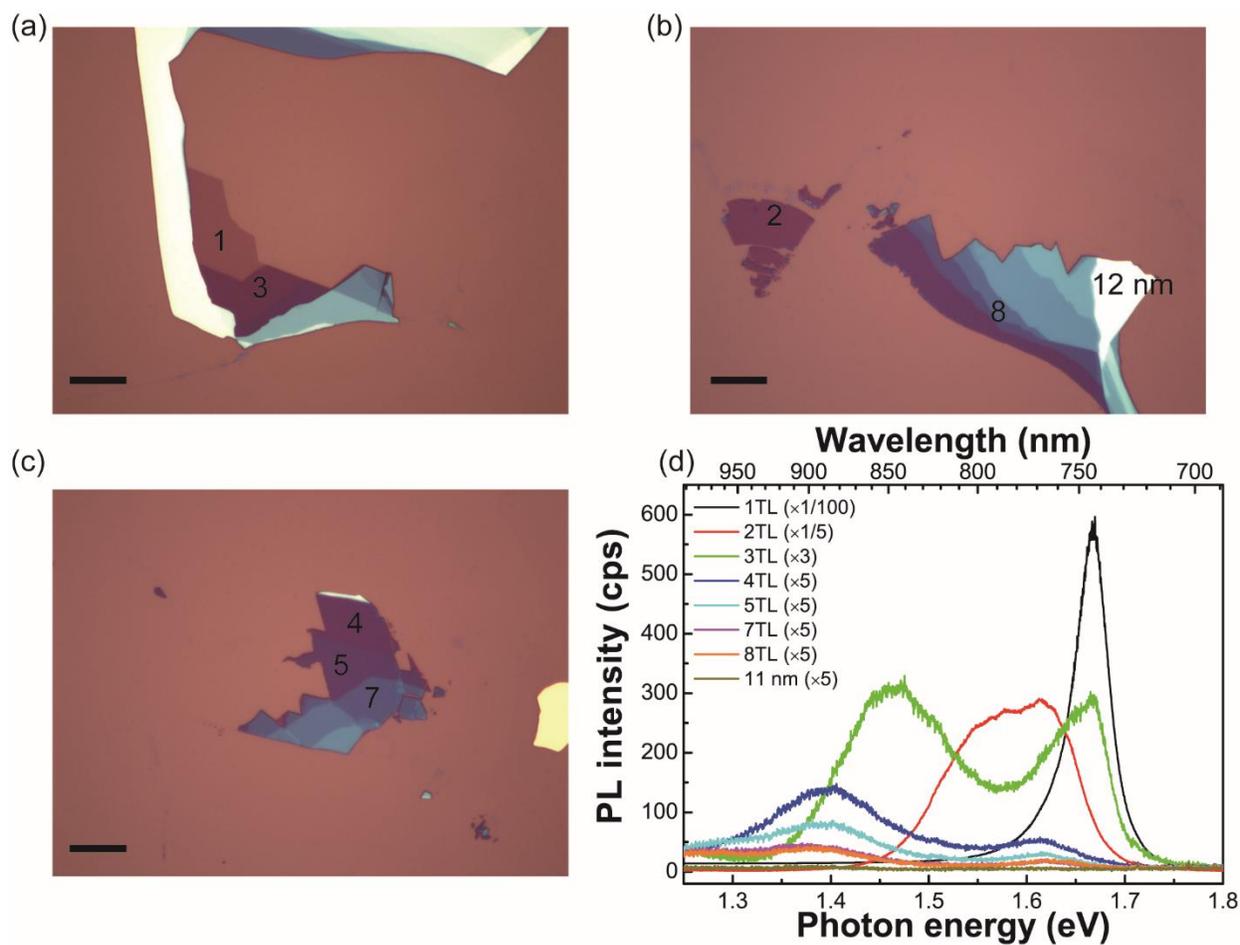

**Figure S1.** (a), (b), (c) Optical microscope images of few-layer WSe$_2$ samples. The scale bar is 10 μm and the number of layers of WSe$_2$ is indicated. (d) Photoluminescence (PL) spectra of few-layer WSe$_2$.

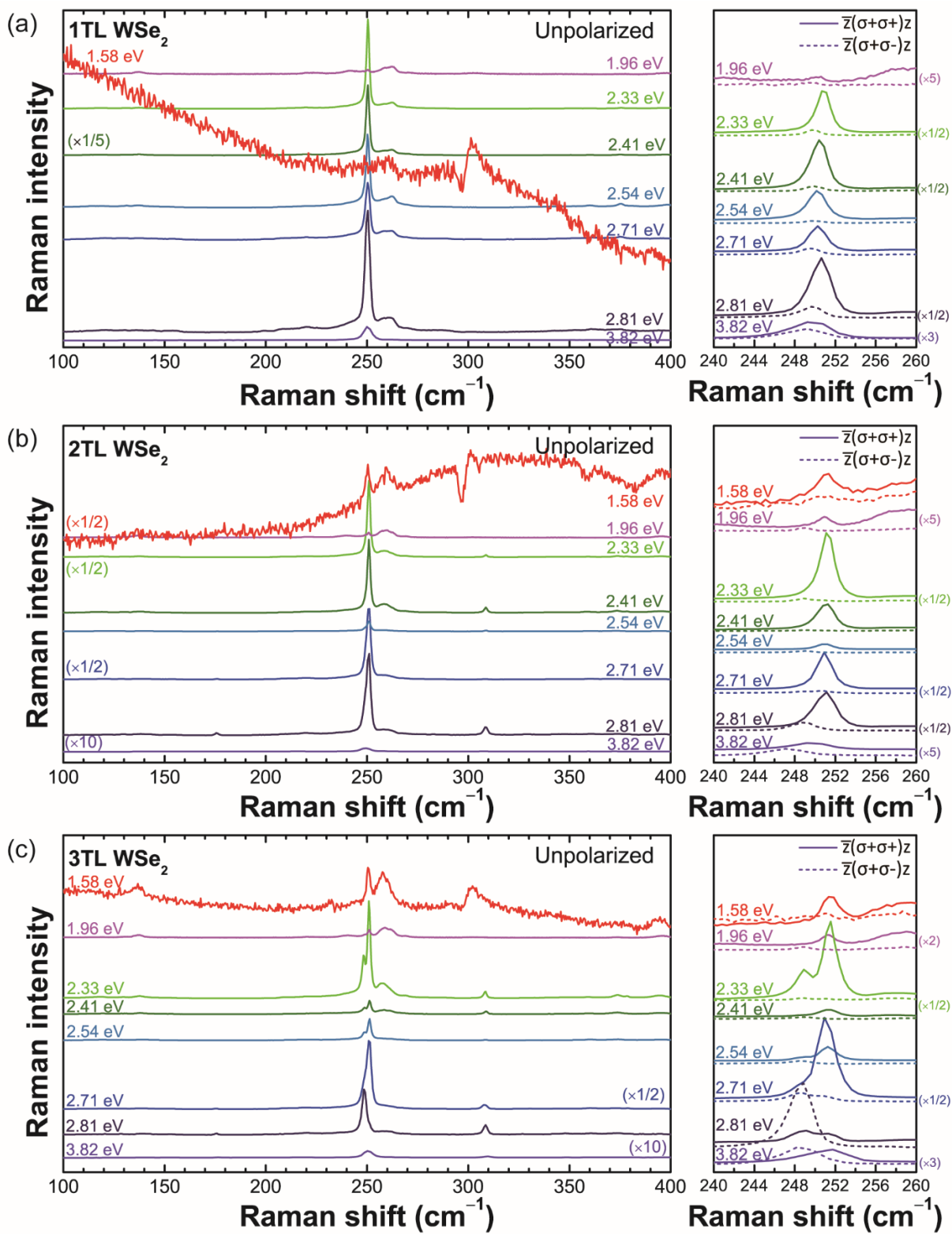

**Figure S2.** Raman spectra of (a) 1TL, (b) 2TL, and (c) 3TL WSe$_2$ measured with 8 excitation energies indicated. Circularly polarized Raman results are shown on the right.

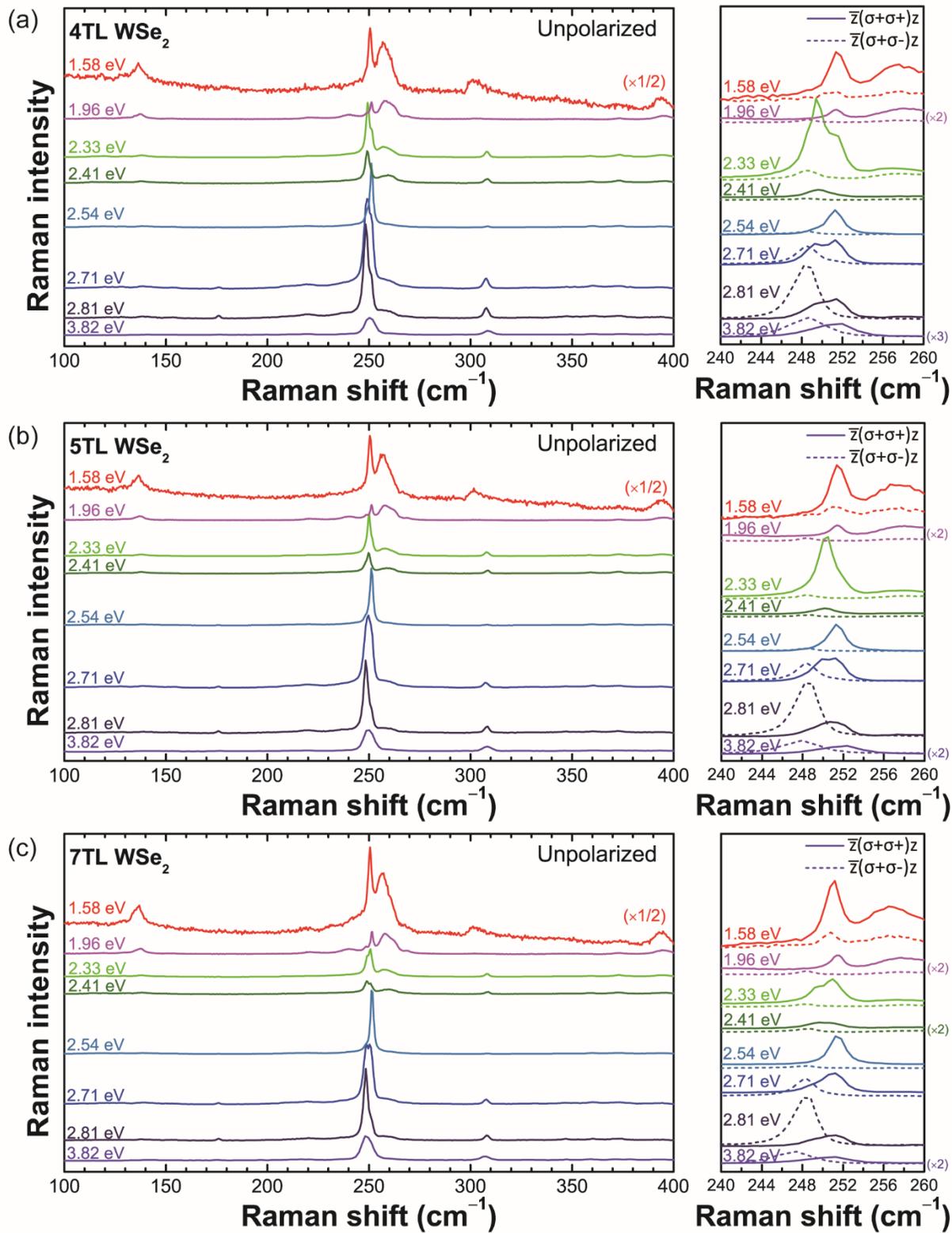

**Figure S3.** Raman spectra of (a) 4TL, (b) 5TL, and (c) 7TL WSe$_2$ measured with 8 excitation energies indicated. Circularly polarized Raman results are shown on the right.

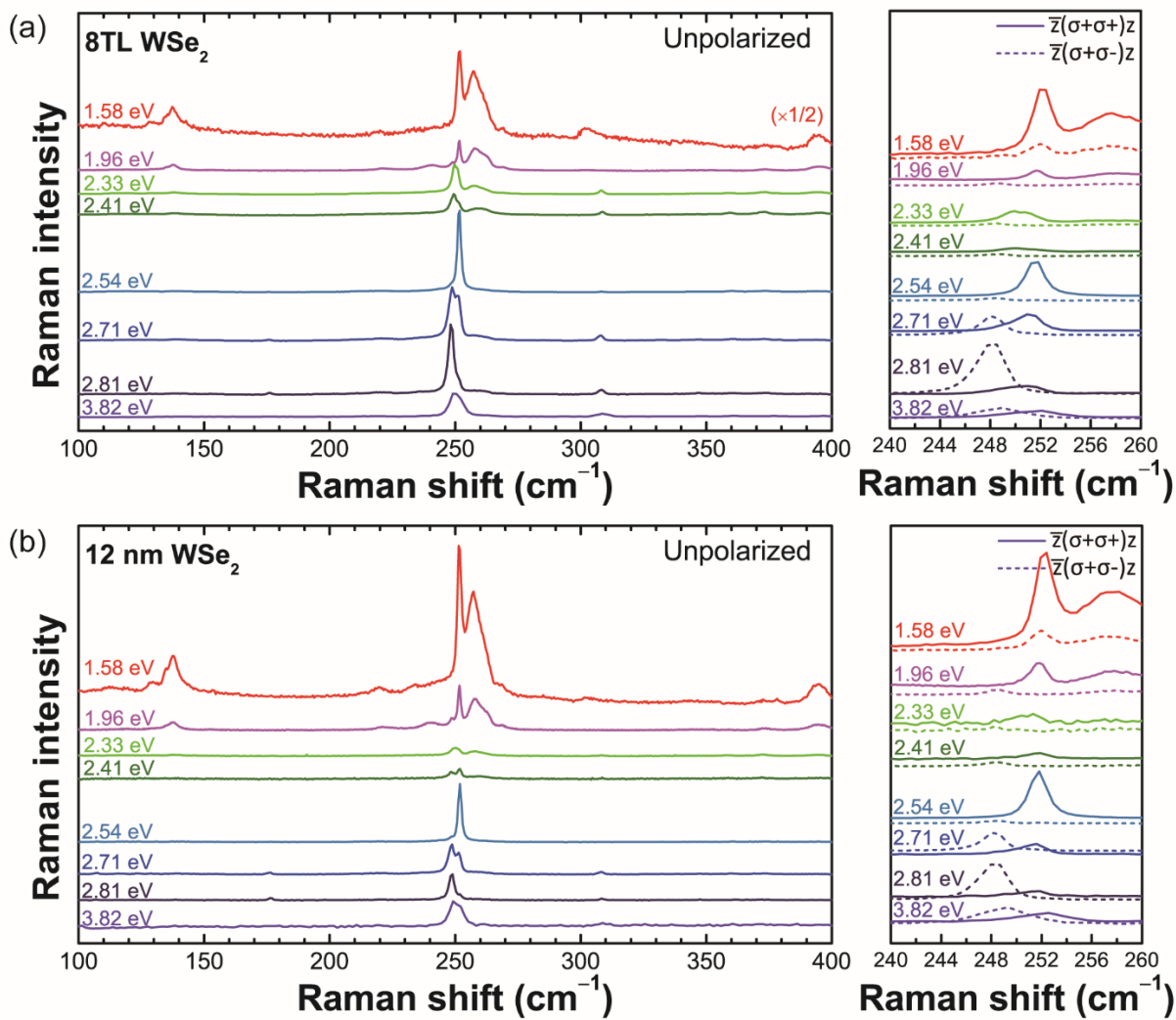

**Figure S4.** Raman spectra of (a) 8TL and (b) 12 nm WSe$_2$ measured with 8 excitation energies indicated. Circularly polarized Raman results are shown on the right.

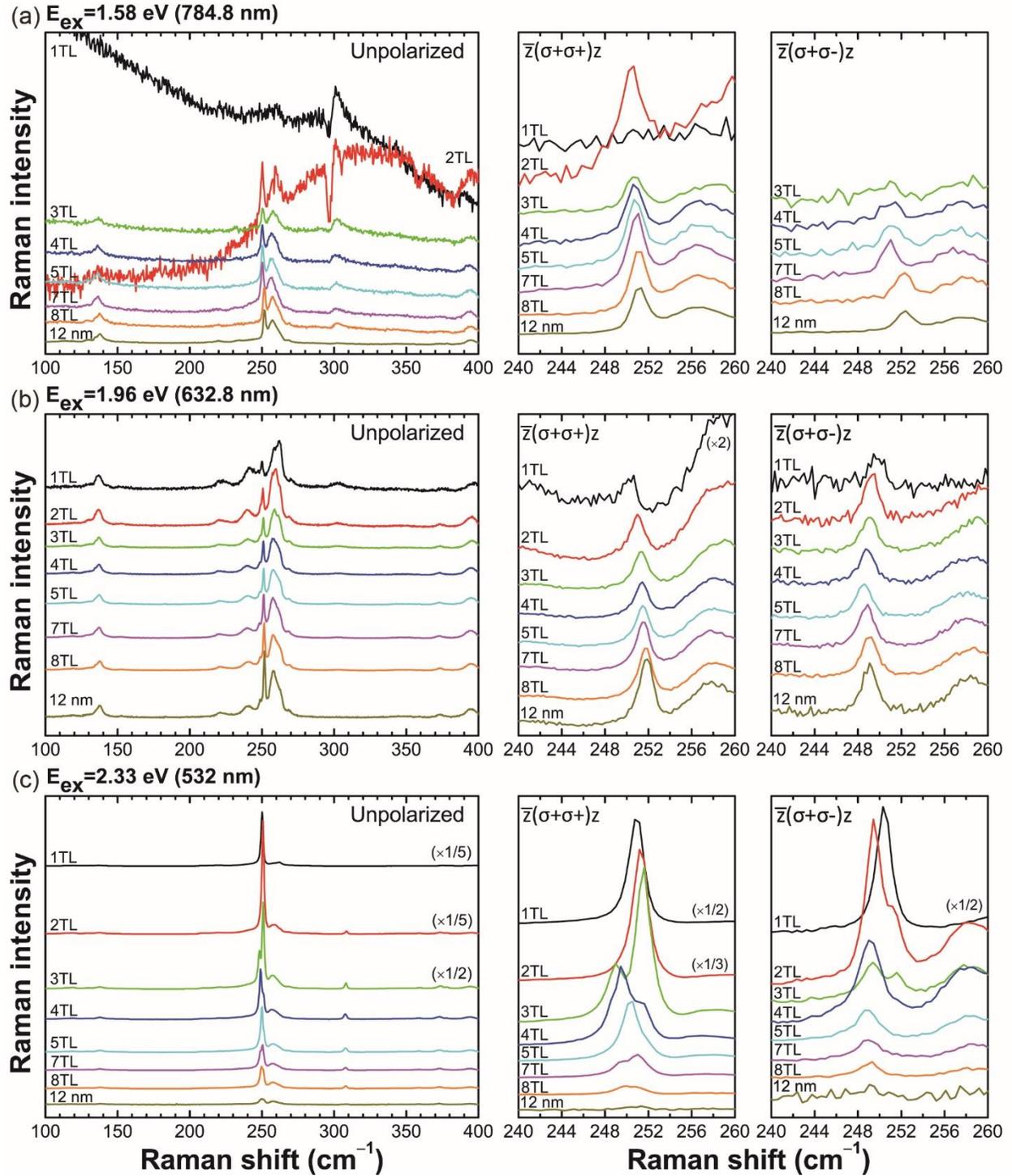

**Figure S5.** Thickness dependence of the Raman spectrum of WSe$_2$ measured with excitation energies (a) 1.58 eV, (b) 1.96 eV, and (c) 2.33 eV. Circularly polarized Raman results are shown on the right.

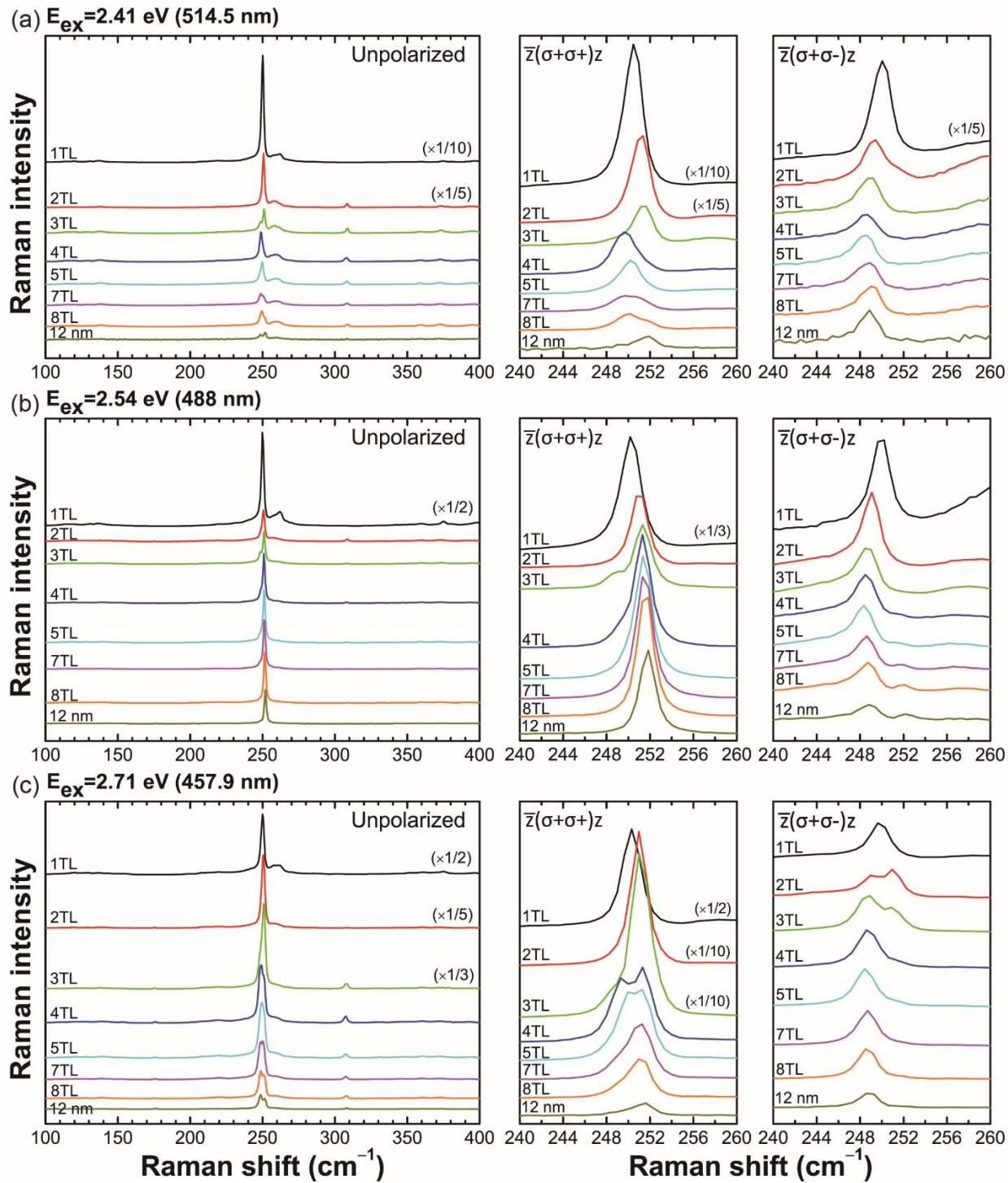

**Figure S6.** Thickness dependence of the Raman spectrum of WSe$_2$ measured with excitation energies (a) 2.41 eV, (b) 2.54 eV, and (c) 2.71 eV. Circularly polarized Raman results are shown on the right.

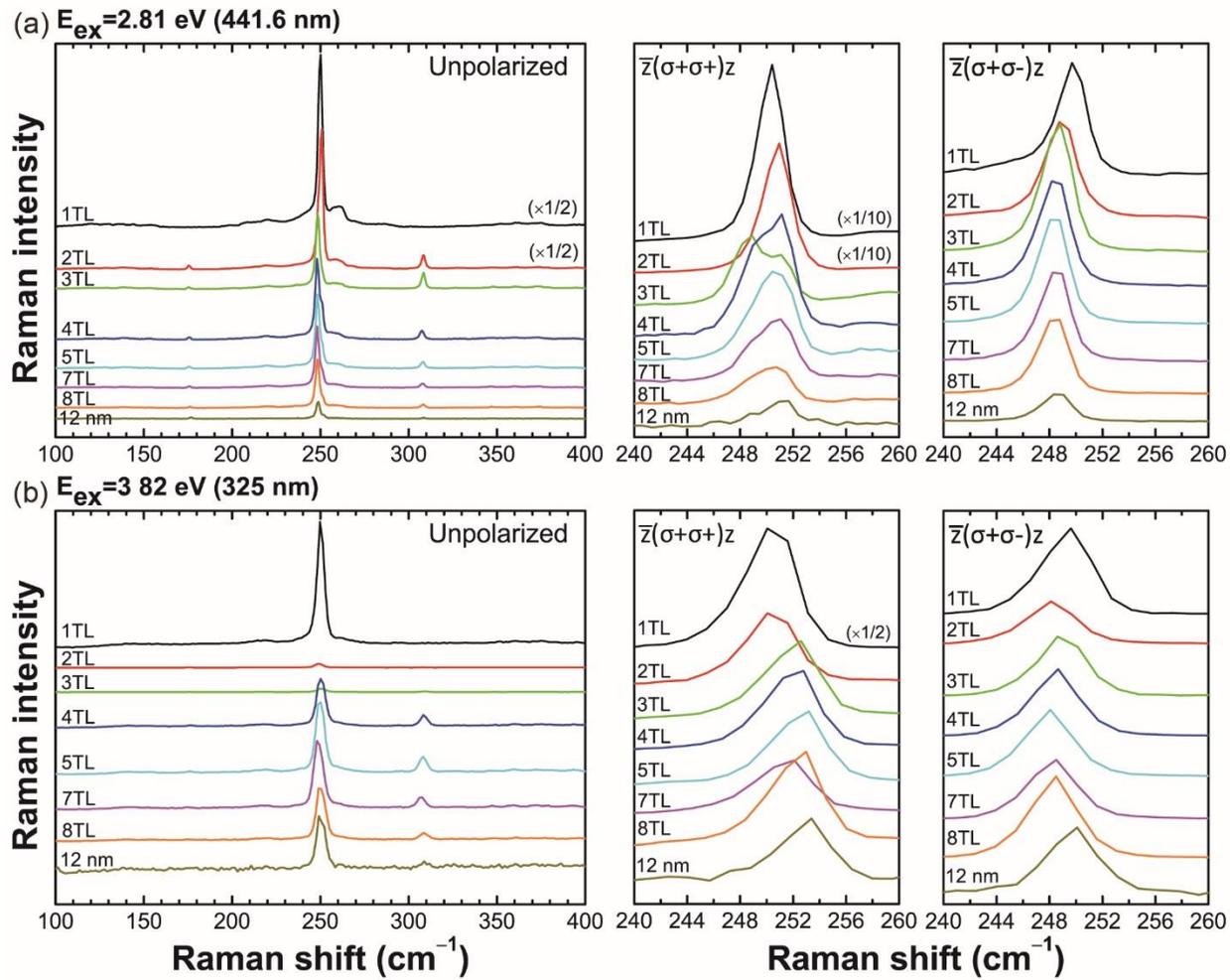

**Figure S7.** Thickness dependence of the Raman spectrum of WSe$_2$ measured with excitation energies (a) 2.81 eV and (b) 3.82 eV. Circularly polarized Raman results are shown on the right.

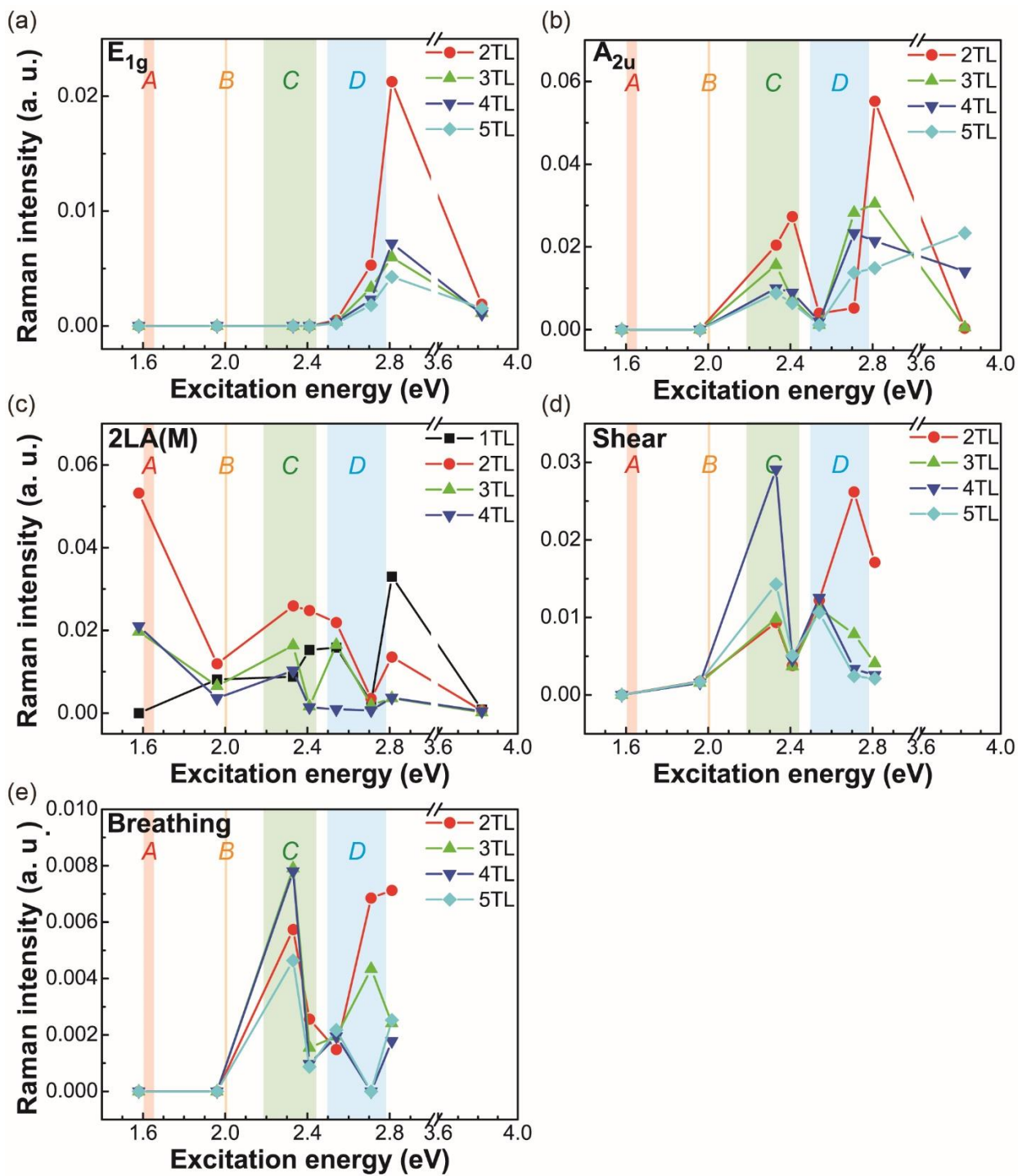

**Figure S8.** Raman intensity as a function of excitation energies for (a) the $E_{1g}$, (b) $A_{2u}$, (c) $2LA(M)$, (d) shear, and (e) breathing modes.

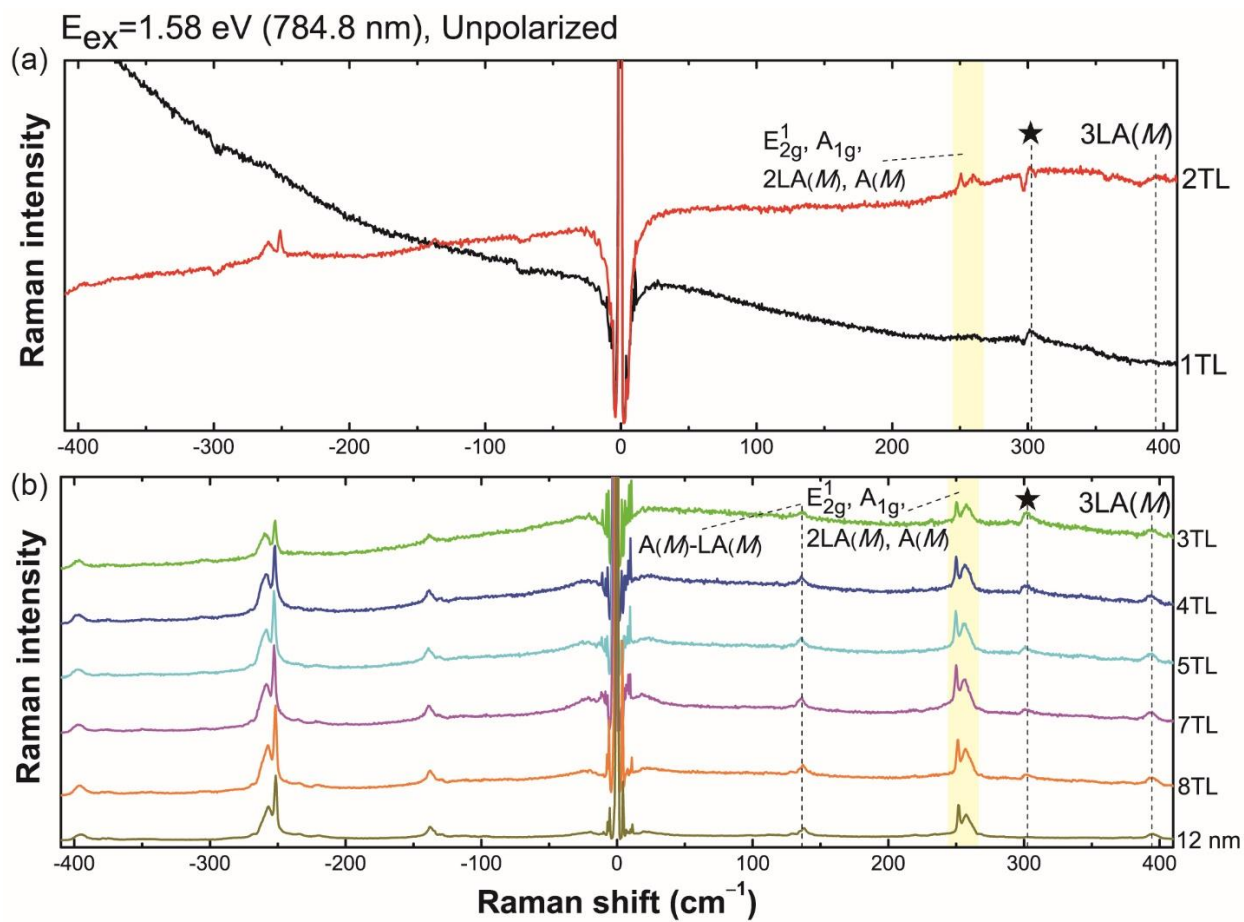

**Figure S9.** Anti-Stokes and Stokes Raman spectra of few-layer WSe$_2$ measured with 1.58 eV excitation energy.

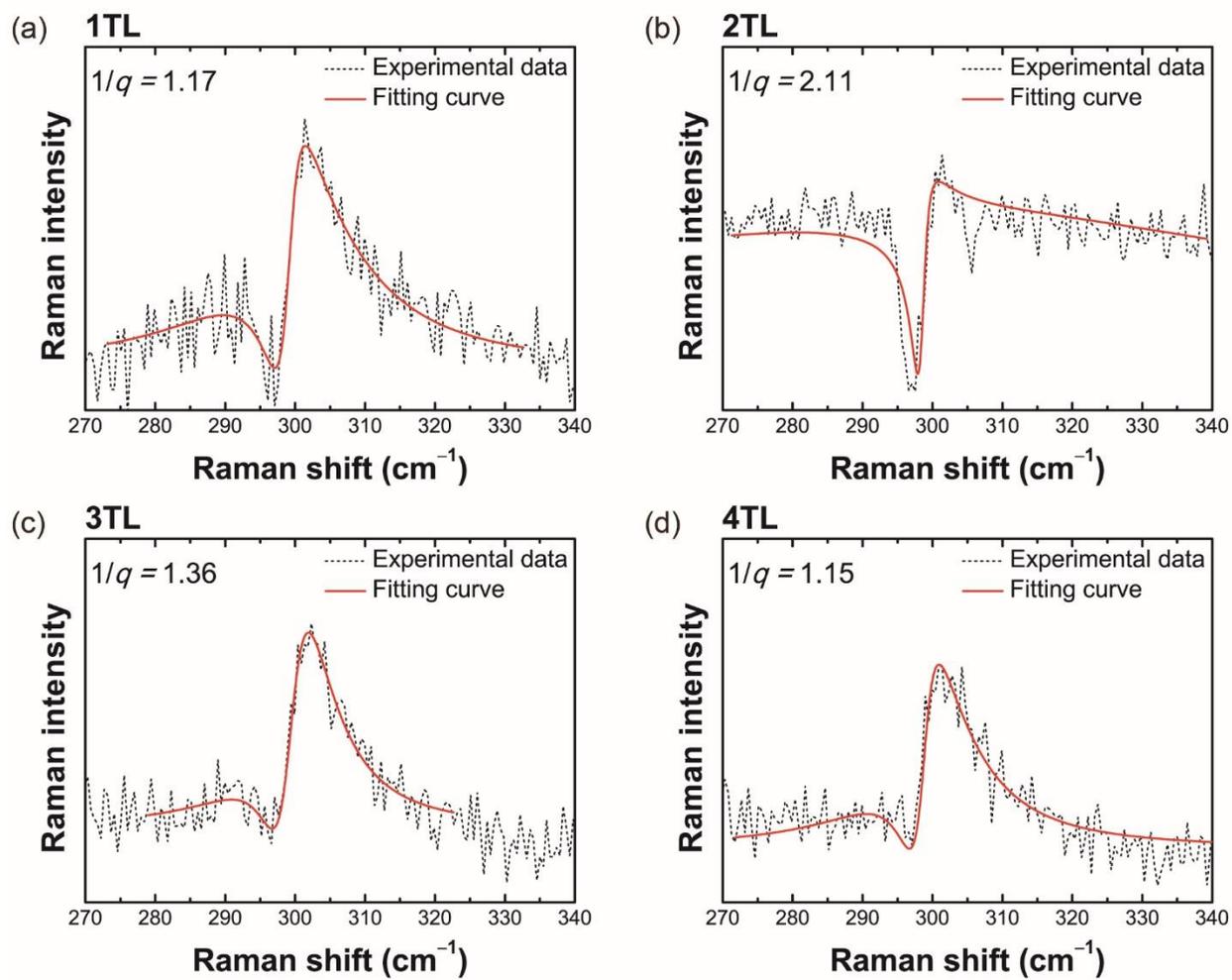

**Figure S10.** Breit-Wigner-Fano (BWF) line fitting of Raman spectra near 301 cm$^{-1}$ measured with excitation energy of 1.58 eV.

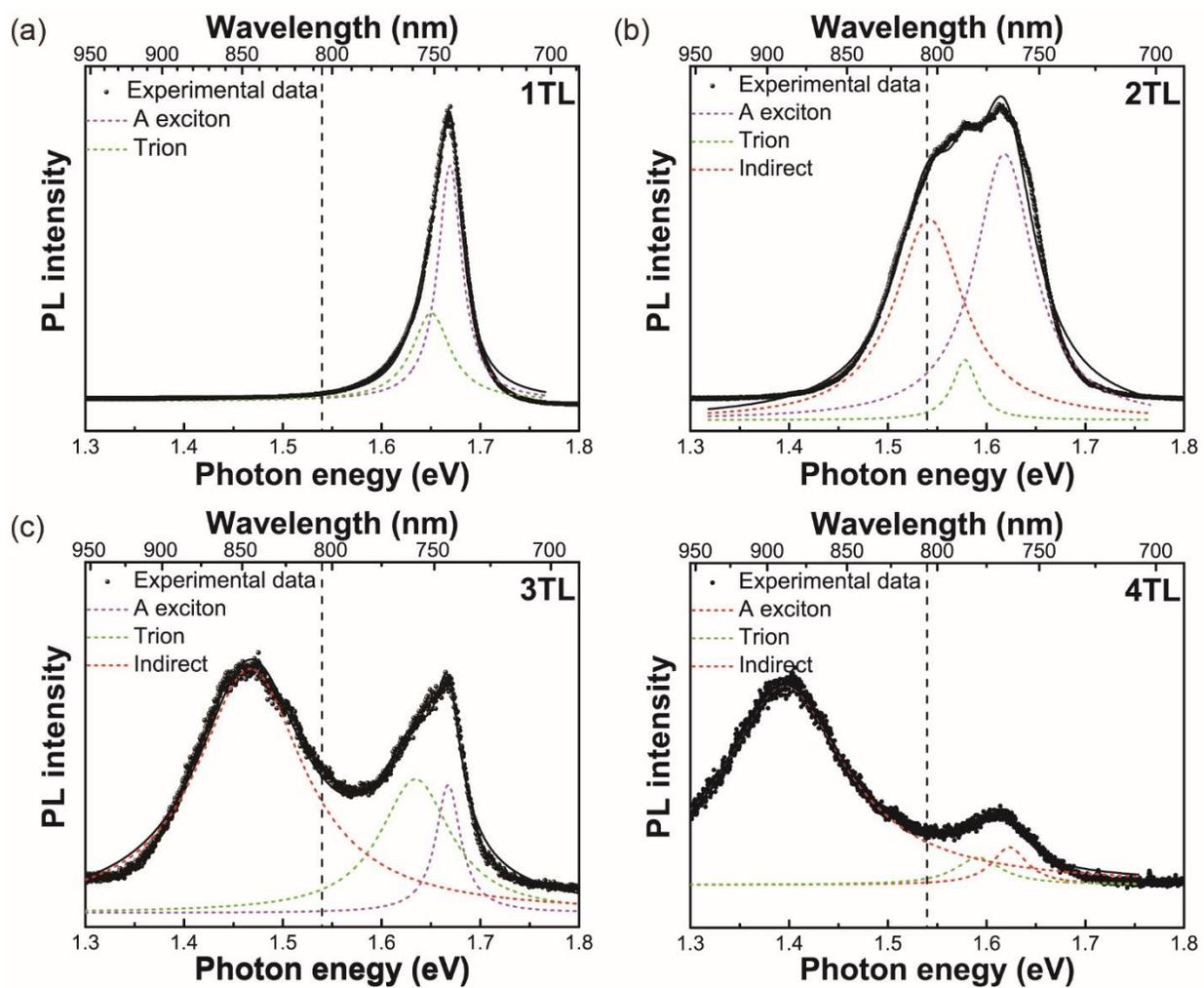

**Figure S11.** Photoluminescence (PL) spectra of 1-4 TL WSe$_2$ with deconvolution of A exciton, trion, and indirect gap transition signals. Black dashed line at 1.54 eV indicates the Raman scattered photon energy of the BWF line at 301 cm$^{-1}$.

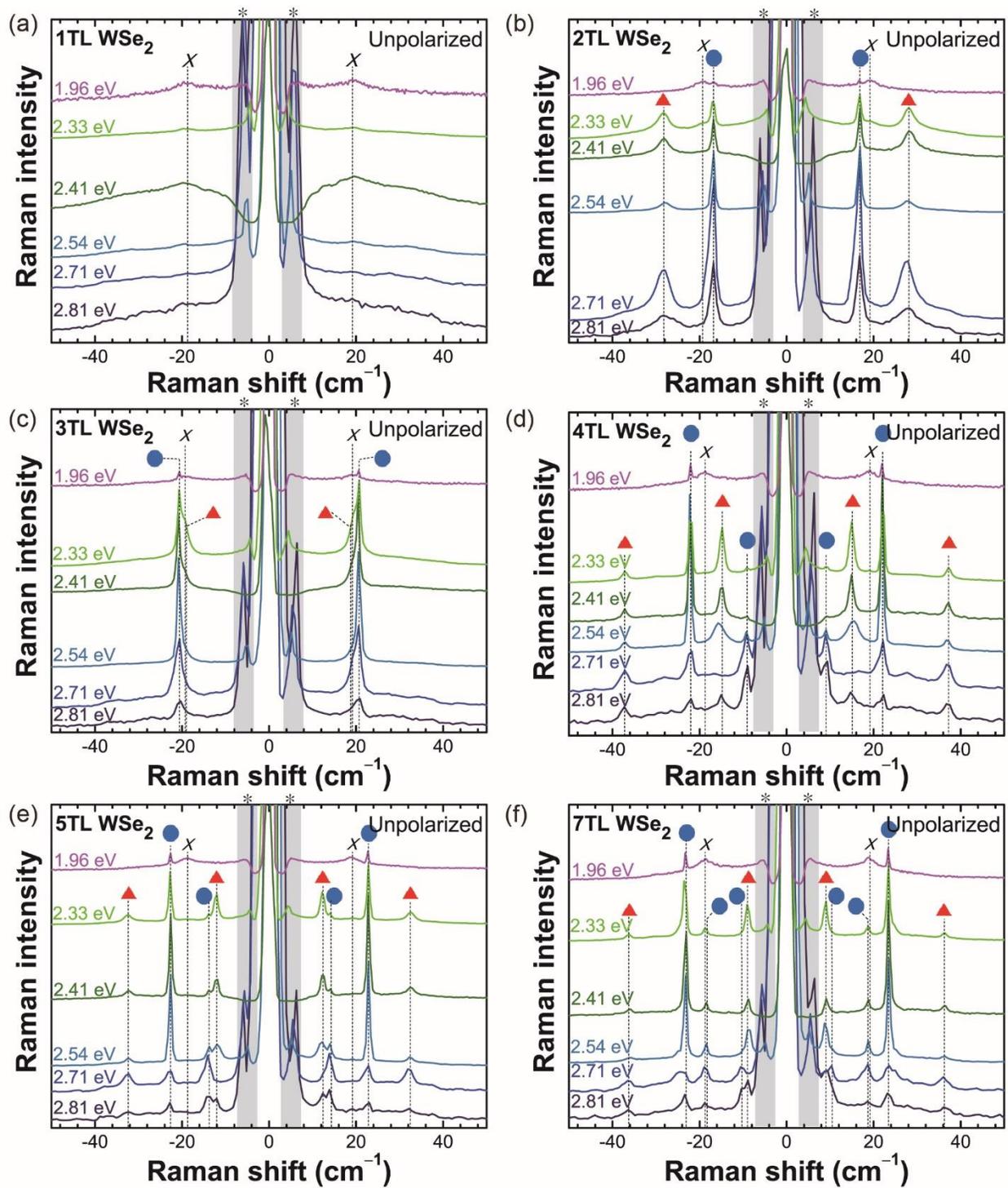

**Figure S12.** Excitation energy dependent low-frequency Raman spectra of 1–5TL and 7TL WSe$_2$. The Brillouin scattering peak of the Si substrate is indicated by (*) and the breathing and shear modes are indicated by red triangles and blue circles, respectively.

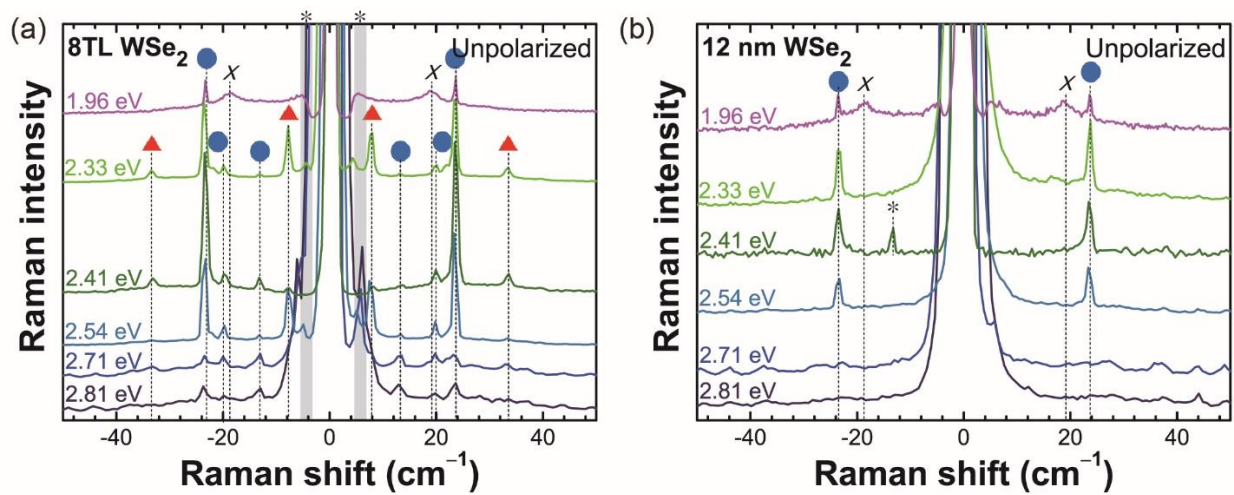

**Figure S13.** Excitation energy dependent low-frequency Raman spectra of (a) 8TL and (b) 12 nm $WSe_2$. A plasma line of the 2.41 eV laser is indicated by (*) and the breathing and shear modes are indicated by red triangles and blue circles, respectively.

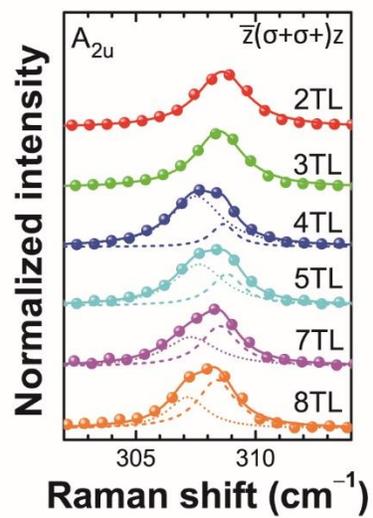

**Figure S14.** Raman spectra of the $A_{2u}$ mode measured with excitation energy of 2.81 eV (441.6 nm).